\RequirePackage{fix-cm}
\documentclass[smallextended]{svjour3}       
\smartqed  
\usepackage{graphicx}
\usepackage[title]{appendix}
\usepackage[version=4]{mhchem}
\usepackage{chemmacros}
    \chemsetup{
        modules = {reactions},
        formula = mhchem,
        reactions/tag-open    = {(R},
        reactions/tag-close   = {)},
    }
\newcommand{\rxnref}[1]{(R\ref{#1})}
\usepackage{comment}
\usepackage{caption}
\usepackage{diagbox}
\usepackage{lineno}
\usepackage{algorithm,algorithmicx,algpseudocode}
\usepackage{amssymb}
\usepackage{float}
\usepackage{here}
\usepackage[square]{natbib}
\usepackage[exponent-product = \cdot]{siunitx}

\begin{document}

\title{Surrogate-based Bayesian Comparison of Computationally Expensive Models: Application to Microbially Induced Calcite Precipitation} 


\titlerunning{Surrogate-based Bayesian Comparison of Computationally Expensive Models: Application to MICP}        

\author{Stefania Scheurer        \and
        Aline Schäfer Rodrigues Silva \and
        Farid Mohammadi \and
        Johannes Hommel \and
        Sergey Oladyshkin \and
        Bernd Flemisch \and
        Wolfgang Nowak
}


\institute{Stefania Scheurer \and Aline Schäfer Rodrigues Silva \and Sergey Oladyshkin \and Wolfgang Nowak \at
              Department of Stochastic Simulation and Safety Research for Hydrosystems (IWS/SimTech), University of Stuttgart, 70569 Stuttgart, Germany \\
              Tel.: +49-711-685 60156\\
              \email{sergey.oladyshkin@iws.uni-stuttgart.de}          
           \and
           Farid Mohammadi \and Johannes Hommel \and Bernd Flemisch \at
          Department of Hydromechanics and Modelling of Hydrosystems (IWS), University of Stuttgart, 70569 Stuttgart, Germany
}

\date{Received: date / Accepted: date}

\maketitle

\begin{abstract}
Geochemical processes in subsurface reservoirs affected by microbial activity change the material properties of porous media. This is a complex biogeochemical process in subsurface reservoirs that currently contains strong conceptual uncertainty. This means, several modeling approaches describing the biogeochemical process are plausible and modelers face the uncertainty of choosing the most appropriate one. The considered models differ in the underlying hypotheses about the process structure. Once observation data becomes available, a rigorous Bayesian model selection accompanied by a Bayesian model justifiability analysis could be employed to choose the most appropriate model, i.e. the one that describes the underlying physical processes best in the light of the available data. However, biogeochemical modeling is computationally very demanding because it conceptualizes different phases, biomass dynamics, geochemistry, precipitation and dissolution in porous media. Therefore, the Bayesian framework cannot be based directly on the full computational models as this would require too many expensive model evaluations. To circumvent this problem, we suggest to perform both Bayesian model selection and justifiability analysis after constructing surrogates for the competing biogeochemical models. Here, we will use the arbitrary polynomial chaos expansion. Considering that surrogate representations are only approximations of the analyzed original models, we account for the approximation error in the Bayesian analysis by introducing novel correction factors for the resulting model weights. Thereby, we extend the Bayesian justifiability analysis and assess model similarities for computationally expensive models. We demonstrate the method on a representative scenario for microbially induced calcite precipitation in a porous medium. Our extension of the justifiability analysis provides a suitable approach for the comparison of computationally demanding models and gives an insight on the necessary amount of data for a reliable model performance.

\end{abstract}

\section{Introduction}
\label{sec:intro}

\subsection{Biogeochemical Processes in Subsurface Porous Media}
\label{sec:intro_biogeochemical}
Biogeochemical processes in porous media are geochemical processes affected by the activity of microbes \citep{lovley1995deep}. They profoundly impact ecosystems as they occur ubiquitously in the subsurface and this makes them interesting for applications in engineering. Some examples of biogeochemical processes that engineers tried to manipulate are: enhanced recovery of resources as in microbially enhanced oil recovery \citep[e.g.][]{bachmann2014biotechnology, mcinerney2005microbially, huang2018experimental}, blocking of preferential flow paths by the accumulation of biomass or minerals precipitated as a result of the microbial metabolism \citep[e.g.][]{bottero2013biofilm, suliman2006change} or bioremediation of soils by microbial decomposition of organic pollutants \citep[e.g.][]{megharaj2011bioremediation, head1998bioremediation, mulligan2003bioremediation}.

However, it is challenging to describe these biogeochemical processes in full detail, because many subprocesses interact in a complex manner \citep{Steefel1996}. Accordingly, it is not easy to control them as desired. A full understanding of these processes is necessary when aiming to control them in order to predict or even regulate the outcome. Thus, modeling is a crucial tool to predict the response of systems under certain conditions \citep{hunter1998kinetic}. Corresponding models are an essential tool in investigating the coupled transport of fluids and reactive substances through porous media and the resulting chemical reactions in the pores \citep{Steefel2005, MacQuarrie2005, Xu2006}.

Several transport models dealing with the biogeochemical process of microbially induced calcite precipitation (MICP) have been discussed in works by e.g. \citet{Martinez2009, Ebigbo2012, hommel2015revised, hommel2016finding, Wijngaarden2016, nassar2018large}. This induced calcite precipitation provides a practical technical application. By accumulating the precipitated calcite, the porosity and permeability of a porous medium can be reduced \citep[e.g.][]{Stocks-Fischer1999, Dupraz2009a, phillips2013potential, Cuthbert2013,Mitchell2013}. Additionally, MICP can be used to reduce erosion or increase soil stability \citep[e.g.][]{Whiffin2007, Gomez2006, Paassen2010, Yang2020}. MICP has been proven to reduce permeability and enhance mechanical strength even at large, field-relevant scales \citep[e.g.][]{Paassen2010, Phillips2016, nassar2018large, Minto2019, Kirkland2020}.

Biogeochemical models are useful, for example, to design, monitor, and evaluate such applications, e.g. to mitigate leakages from a geological gas reservoir into above aquifers in advance \citep[e.g.][]{Cuthbert2013, nassar2018large, cunningham2019field, Minto2019, landa2020practical}. Our limited knowledge about the interaction of the processes that govern biogeochemical systems leads to several modeling approaches that differ, e.g., in their level of detail. The uncertainty of choosing between these modeling alternatives is known as conceptual uncertainty.

\subsection{Conceptual Uncertainty}
\label{sec:intro_conceptual}
When modeling an environmental process, we have to make assumptions and simplifications because, usually, the real process is too complex to be represented in full detail. Consequently, one has to deal with various types of uncertainty. Besides input and parameter uncertainty, conceptual uncertainty (uncertainty of model choice) has to be taken into account. If we chose a single model and did not consider possible alternatives, we might strongly underestimate the overall prediction uncertainty because the space of potential models is not sufficiently covered \citep{enemark2019hydrogeological, refsgaard2012review, rojas_conceptual_2008}.

Many studies have identified conceptual uncertainty as a key source of uncertainty in modeling \citep[e.g.][]{burnham2002model, neuman2003maximum, hojberg_model_2005, rojas_conceptual_2008, rojas2010application, gupta2012towards, troldborg2007importance, refsgaard2012review, renard_understanding_2010, schoniger2015statistical, enemark2019hydrogeological}. These studies suggest to treat modeling concepts with different levels of detail and different assumptions as competing hypotheses. By using statistical techniques such as Bayesian model selection (BMS), we can evaluate which model is the most appropriate representation of the system \citep{raftery_bayesian_1995, wasserman_bayesian_2000}.

However, two challenges persist. First, it is important to note that there is no existing method which allows to quantify conceptual uncertainty on an absolute level \citep{nearing2018ensembles, hoge2019hydrologist}. Second, biogeochemical modeling, discussed briefly in Section \ref{sec:intro_biogeochemical}, is computationally very demanding since it conceptualizes different processes in subsurface porous media. Thus, a direct application of the rigorous probabilistic machinery is not feasible due to a necessity of a high number of model evaluations. In this study, we address the second challenge.

\subsection{Surrogate Representation of the Underlying Physical Models}
\label{sec:intro_pce}
In order to assure feasibility of the probabilistic BMS framework, we will construct surrogate models for each corresponding version of the biogeochemical model. The main goal of a surrogate model is to replicate the behavior of the underlying physical model from a limited set of runs without sacrificing a lot of detail and accuracy. For constructing a surrogate the original model should be evaluated by using those sets of modeling parameters out of various possibilities that covers the parametric space as good as possible. Considering very high computational costs of biogeochemical models, whereby one model evaluation requires days, we need to select an approach that will capture the main features of the underlying physical models after a very small number of model evaluations. Following a recent benchmark comparison study by \citep{Koeppel2019}, we construct the surrogate model using the arbitrary polynomial chaos expansion technique (aPC) introduced in \citep{oladyshkin1}, which is suitable for our purpose.

In short, the data-driven aPC approach can be seen as a machine learning approach that approximates the model output by its dependence on model parameters via multivariate polynomials. The data-driven feature of aPC offers complete flexibility in the choice and representation of probability distributions. It requires no approximation of a density function, that usually caused additional uncertainties \citep{oladyshkin2}. Based on the original polynomial chaos expansion introduced by \citep{Wiener1938}, the aPC constructs surrogate models with the help of an orthonormal polynomial basis. Such a reduction of a full biogeochemical model into a surrogate model offers the path to perform a rigorous stochastic analysis at strongly reduced computational costs.

\subsection{Two-Stage Bayesian Model Selection Procedure}
\label{sec:intro_bayesian}
Bayesian model selection \citep[e.g.][]{raftery_bayesian_1995, wasserman_bayesian_2000} has been used in many fields of research to support the choice between competing models \citep[e.g.][]{mohammadi2018bayesian, wohling_bayesian_2015, schoniger2015finding, hooten_guide_2015, Parkinson_Bayesian_2006, cremers_stock_2002, brunetti2020handling}. It ranks competing physical models based on their quality to represent the available measurement data. To be more specific, BMS employs the Bayesian model evidence (BME) as the score indicating the quality of the model against the available data.

Here, we will consider several models describing biogeochemical processes in subsurface porous media. They contain various assumptions helping to simplify the modeling procedure. The BME-based ranking follows the principle of parsimony \citep{schoniger2014model} or rather ``Occam's razor'', which tells to ``choose the simplest one between competing hypotheses ''\citep{jefferys_ockhams_1992}. The work by \citet{schoniger2015finding} uses this property to find a justifiable level of complexity (variability of the model) for modeling a certain quantity of interest.

Following the framework introduced in \citet{schoniger2015finding}, we will adopt a two-stage approach for model testing. In the first stage, the classical BMS procedure is used, in which models are tested against measurement data. This procedure is complemented by the second stage, the so-called model justifiability analysis. Here, competing models are tested against each other based on ``synthetic true'' data instead of measurement data. Based on this analysis, one can diagnose similarities between competing models and identify a suitable model that is ``affordable'' when only a realistic amount of measurement data is given. A joint interpretation of both stages provides insights that help to find the most appropriate model, which represents the observed system best under acceptable computational costs.

\subsection{Goals and Paper Structure}
\label{sec:intro_aim}
The overall aim of this study is to set up a rigorous ranking of biogeochemical computationally expensive models introducing the surrogate-based two-stage Bayesian model selection procedure. We extend the Bayesian justifiability analysis introduced by \citet{schoniger2015finding}. Our novel correction factor allows the use of surrogate models and thus, makes this analysis suitable for computationally demanding models.

Section \ref{sec:methods} introduces necessary details on Bayesian updating of the aPC expansion and extends the Bayesian model selection of computationally demanding models to the Bayesian model justifiability analysis introducing novel correction factors. Section \ref{sec:models} introduces the biogeochemical process of microbially induced calcite precipitation (MICP) and the corresponding model set. Section \ref{sec:results} performs Bayesian model selection among MICP models and assesses their similarity using the novel surrogate-based justifiability analysis. Section \ref{sec:summary} summarizes the results and gives an outlook for further investigation.

\section{Bayesian Assessment of Computationally Demanding Models}
\label{sec:methods}

\subsection{Arbitrary Polynomial Chaos Expansion}
\label{sec:methods_pce}
We will consider computationally demanding models, for which a straightforward application of the Bayesian model selection procedure is infeasible. Therefore, we will construct so-called surrogate models with negligible computational costs to replicate the behavior of the original physical models via the arbitrary polynomial chaos expansion (aPC) approach introduced in \citep{oladyshkin1}. Surrogate models are mapping the modeling parameters to the model output, capturing the main features of the underlying physical model. In what follows, we present the core idea for the construction of these aPC-based surrogate models.

Let $\boldsymbol{\omega} = (\omega_1,...,\omega_{N_p})$ represent the $N_p$-dimensional vector of model parameters, whereby all parameters in $\boldsymbol{\omega}$ are assumed to be independent \citep{oladyshkin1}. Let the model responses be given in the form of $M_k = f(\boldsymbol{x},t; \boldsymbol{\omega})$, where $M_k$ can be some differential equation, a coupled system of differential equations or just a simple function. Moreover, the model parameters can depend on a certain point in space $\boldsymbol{x} = (x_1,x_2,x_3)$ and time $t$. The model responses $M_k$ can be approximated with a spectral projection of responses onto orthogonal polynomial bases as follows:

\begin{align}
    M_k(\boldsymbol{x},t; \boldsymbol{\omega}) \approx \tilde{M}_k(\boldsymbol{x},t; \boldsymbol{\omega}) =  \sum\limits_{s=0}^{D} c_s(\boldsymbol{x},t) \cdot \Psi _s(\boldsymbol{\omega}) \text{,}
    \label{eq:modelapprox}
\end{align}

with polynomials $\Psi _s(\boldsymbol{\omega})$ of the multivariate orthogonal polynomial basis. These polynomials are constructed according to \citet{oladyshkin2}. There are $D$ polynomials needed for the expansion, whereby $D$ is the number of expansion coefficients dependent on the number of model parameters $N_p$ and the chosen maximum polynomial degree $d$: $\displaystyle D = (N_p+d)!/(N_p!d!)-1$. The coefficients $c_s(\boldsymbol{x},t)$ depend on space and time since the original model output depends on space and time.

To compute the coefficients $c_s(\boldsymbol{x},t)$ of the polynomial chaos expansion in equation (\ref{eq:modelapprox}), we employ a non-intrusive stochastic collocation method \citep{oladyshkin1}. The non-intrusiveness of this method implies that the model $M_k$ can be considered as a black box, so that there is no need of making changes in the governing equations of the original model at hand. Using this method, a finite number of model evaluations $D$ is sufficient to determine the coefficients. The coefficients (using the $D$ evaluations of model $M_k$ on $D$ so-called collocation points $\left\{\omega_1^{(i)},...,\omega_{N_p}^{(i)}\right\}$, $i = 1,...,D$) can be computed by the following system of equations:

\begin{align}
    \begin{bmatrix}
        \Psi_1\left(\boldsymbol{\omega}^{(0)}\right) & ... & \Psi_D\left(\boldsymbol{\omega}^{(0)}\right) \\
        ... & ... & ... \\
        \Psi_1\left(\boldsymbol{\omega}^{(D)}\right) & ... & \Psi_D\left(\boldsymbol{\omega}^{(D)}\right)
    \end{bmatrix}
    \cdot
    \begin{bmatrix}
        c_0(\boldsymbol{x},t) \\
        ... \\
        c_D(\boldsymbol{x},t)
    \end{bmatrix}
    =
    \begin{bmatrix}
        M_k\left(\boldsymbol{x},t; \boldsymbol{\omega}^{(0)}\right) \\
        ... \\
        M_k\left(\boldsymbol{x},t; \boldsymbol{\omega}^{(D)}\right)
    \end{bmatrix}
    \label{eq:CoeffLGS}
\end{align}

or

\begin{align}
    \boldsymbol{\Psi}(\boldsymbol{\omega}) \cdot \textbf{c}(\boldsymbol{x},t) = \boldsymbol{M_k}(\boldsymbol{x},t;\boldsymbol{\omega})\text{.}
    \label{eq:sCoeffLGS}
\end{align}

The $D \times D$ matrix $\boldsymbol{\Psi}$ contains the basis polynomials, evaluated on different collocation points, and the vector $\textbf{c}$ of size $D \times 1$ contains the expansion coefficients. The outputs of the model $M_k$ on the different collocation points are represented by vector $\boldsymbol{M_k}$ of size $D \times 1$. If one aims to compute the surrogate model of $M_k$ for different points in time, it is sufficient to compute the matrix $\boldsymbol{\Psi}$ once for a fixed amount of parameters and collocation points and an expansion degree $D$, since the matrix is space and time independent, unlike both of the vectors $\textbf{c}$ and $\boldsymbol{M_k}$. Accordingly, the coefficients are computed based on the model output using the collocation points for different points in space and time separately (available Matlab code in \citet{aPC_Matlab}). 

The solution of the system of equations \eqref{eq:sCoeffLGS} is obviously dependent on the choice of the collocation points $\left\{\omega_1^{(i)},...,\omega_{N_p}^{(i)}\right\}$, $i = 1,...,D$. According to \citet{villadsen1978solution} the optimal collocation points are the roots of the univariate polynomials used for the construction of the multivariate polynomial basis of degree $D+1$ \citep{oladyshkin1}.

Hence, the resulting surrogate model represents the original model at the collocation points exactly while some ``polynomial interpolation'' is applied between them or rather an extrapolation outside of the range of the collocation points \citep{mohammadi2018bayesian}.

\subsection{Bayesian Updating of the aPC-Based Surrogate Representation}
\label{sec:methods_iterative}

The procedure described in Section \ref{sec:methods_pce} can be seen as an initial step, whereby the surrogate representation of the original model makes use of the prior distribution of the modeling parameters and omits the available measurement data. Therefore, the constructed surrogate model $\tilde{M}_k$ could be imprecise and may not necessarily cover well the region of the parameter space where the measurement data is relevant (i.e. posterior). Using a higher expansion degree to improve the surrogate model globally would increase the computational time excessively.

Therefore, to overcome this issue, we employ an iterative Bayesian updating process of the aPC representation (BaPC) that improves the accuracy of the surrogate by incorporating new collocation points at approximate locations of the maximum a posteriori parameter set \citep{oladyshkin2013bayesian}. The idea is to evaluate the surrogate model $\tilde{M}_k$ on a high number of parameter realizations, obtained from their prior distribution, to weigh the points by their posterior probability. As the parameter realization with the highest posterior probability is assumed to be in the parameter region of interest, the surrogate model should be refined there. According to the BaPC strategy, we will evaluate the original model $M_k(\boldsymbol{x},t;\boldsymbol{\omega})$ on the suggested new collocation point $\boldsymbol{\omega}$ corresponding to the maximum a posteriori parameter set and recalculate the expansion coefficients $\textbf{c}(\boldsymbol{x},t)$ by solving equation (\ref{eq:sCoeffLGS}). The increasing number of collocation points leads to an overdetermined system of equations for the determination of the coefficients as described in Appendix \ref{app:system}. In this way, we iteratively update the aPC representation in equation (\ref{eq:modelapprox}) by incorporating the points where the probability to capture the measurement data is higher. This process is repeated until the surrogate model captures the measurement data sufficiently well, although the number of iterations should be limited to keep the computational costs manageable (Matlab code available in \citet{BaPC_Matlab}).

The suggested BaPC framework has shown promising results for computationally very demanding models \citep[e.g.][]{oladyshkin2013chaos, mohammadi2018bayesian, beckers2020bayesian} and further details are shown in \citet{oladyshkin2013bayesian}. Alternatively, other Bayesian strategies can be found in \citet{oladyshkin2020bayesian3}.

\subsection{Approximation Quality of aPC-based Surrogate Models}
\label{sec:methods_quality}

To assess the quality of a constructed surrogate model during the iterative Bayesian updating of an aPC expansion, we will estimate the approximation error in equation (\ref{eq:modelapprox}). Since the stochastic collocation belongs to the family of regression methods, only calculating the error at the collocation points would lead to biased results. Yet, computing the validation error via so-called testing parameter sets to assess the accuracy of the model, trained on the training collocation points, is computationally infeasible.

To remedy this problem, one can use the leave-one-out cross validation (LOOCV) as described in \citet{blatman2010adaptive} instead. The collocation points are divided $P$ times into two subsets, assuming that the set of collocation points is of size $P \geq D+1$: for the calculation of the coefficients the collocation points are omitted one after the other. After the coefficients have been determined with the help of the remaining collocation points, the resulting surrogate model is evaluated on the omitted collocation point. Then, the difference to $M_k$, evaluated on this point, is computed \citep{blatman2010adaptive}. This is done for all collocation points and finally the mean value over all quadratic errors is determined:

\begin{align}
    \overline{err}_{\text{LOOCV}} = \frac{1}{P} \cdot \sum\limits_{i=1}^P \left(M_k\left(\boldsymbol{\omega}^{(i)}\right) - \tilde{M_k}_{\backslash\boldsymbol{\omega}^{(i)}}\left(\boldsymbol{\omega}^{(i)}\right)\right)^2 \text{,}
    \label{eq:MittelLOOCV}
\end{align}

where $P$ is the current number of collocation points, $M_k\left(\boldsymbol{\omega}^{(i)}\right)$ is the model evaluated on the omitted collocation point $\boldsymbol{\omega}^{(i)}$ and $\tilde{M_k}_{\backslash \boldsymbol{\omega}^{(i)}}\left(\boldsymbol{\omega}^{(i)}\right)$ is the surrogate model constructed without the collocation point $\boldsymbol{\omega}^{(i)}$ evaluated on the collocation point $\boldsymbol{\omega}^{(i)}$.

\subsection{Bayesian Model Selection}
\label{sec:methods_bayesian}
 Bayesian Model Selection allows to rank models based on their probability to be the data-generating process \citep[e.g.][]{raftery_bayesian_1995, wasserman_bayesian_2000, hoge2019hydrologist}. For this ranking, prior model weights $P\left( M_k \right)$ are updated to posterior model weights $ P(M_k|\boldsymbol{y}_0)$ using Bayes' theorem:
 
\begin{align}
    P(M_k|\boldsymbol{y}_0) = \frac{p(\boldsymbol{y}_0|M_k) P(M_k)}{\sum_{i=1}^{N_\text{m}} p(\boldsymbol{y}_0|M_i) P(M_i)} \text{,}
    \label{eq:bayes}
\end{align}

with $\boldsymbol{y}_0$ being the vector of measurements and the models' prior probability $P(M_k)$. The prior probability $P(M_k)$ is a subjective estimation of the investigator or the modeler about which model is an exact representation of the data-generating process, without actually knowing the data yet \citep{raftery_bayesian_1995}. Uniformly distributed priors $P(M_k) = \frac{1}{N_\text{m}}$ with $N_\text{m}$ competing models are a common choice. The term $p(\boldsymbol{y}_0|M_k)$ is the so-called Bayesian Model Evidence (BME). The BME value is also known as marginal likelihood, because it can be calculated by averaging (marginalizing) over the parameter space $\Omega_k$ of each model \citep{kass_bayes_1995, schoniger2014model}. The marginalization makes BME independent of the parameter choice and hence it is a characteristic of only the model $M_k$. Accordingly, BME is defined as

\begin{align}
    p(\boldsymbol{y}_0|M_k) = \int_\Omega p(\boldsymbol{y}_0 | M_k, \boldsymbol{\omega})~p(\boldsymbol{\omega} | M_k)~d\boldsymbol{\omega} \text{,}
    \label{eq:bme}
\end{align}

where $p(\boldsymbol{\omega} | M_k)$ is the model-specific prior distribution of the model parameter vector $\boldsymbol{\omega}$. The likelihood function $p(\boldsymbol{y}_0 | M_k, \boldsymbol{\omega})$ quantifies how well the predictions $\boldsymbol{y}_k$ fit the measurement data $\boldsymbol{y}_0$ and includes assumptions on the measurement error \citep{raftery_bayesian_1995}. Here, we will choose a Gaussian likelihood function with zero mean:

\begin{align}
 p(\boldsymbol{y}_0 | M_k, \boldsymbol{\omega}) &= (2\pi)^{-N_\text{s}/2}|\boldsymbol{R}|^{-1/2} \nonumber
 \\
 &\cdot \exp \left( -\frac{1}{2} (\boldsymbol{y}_0 - \boldsymbol{y}_k(\boldsymbol{\omega}))^T \boldsymbol{R}^{-1} (\boldsymbol{y}_0 - \boldsymbol{y}_k(\boldsymbol{\omega})) \right) \text{,}
 \label{eq:likelihood}
\end{align}

where $\boldsymbol{R}$ is the covariance matrix of the measurement error $\epsilon$ of size $N_\text{s} \times N_\text{s}$ (with data set size $N_\text{s}$), and $\boldsymbol{y}_k(\boldsymbol{\omega})$ is the prediction made by model $M_k$ with the model parameter vector $\boldsymbol{\omega}$. 

For most applications, there is no analytical solution of equation \eqref{eq:bme} and the corresponding integral could be estimated using a brute-force Monte Carlo approach. To perform the Monte Carlo integration, we create a sample set of $N_{\text{MC}}$ realizations of the modeling parameter vector $\boldsymbol{\omega}$ based on its prior distribution $p(\boldsymbol{\omega} | M_k)$. With the corresponding likelihood functions \eqref{eq:likelihood}, we will obtain the following numerical approximation of the BME value:

\begin{align}
    p(\boldsymbol{y}_0|M_k) \approx \frac{1}{N_{\text{MC}}} \sum\limits_{i=1}^{N_{\text{MC}}} p(\boldsymbol{y}_0|M_k, \boldsymbol{\omega}_i) \text{,}
    \label{eq:MC}
\end{align}

where $\boldsymbol{\omega}_i$ is the $i$-th parameter realization for model $M_k$.

\subsection{aPC-Based Bayesian Model Selection}
\label{sec:methods_bayesian_pce}

Remarking that the surrogate representation $\tilde{M}_k$ is only an approximation of the original model $M_k$, we expect that surrogate-based BME values could be misleading for the Bayesian model selection procedure. Therefore, conclusions drawn from BME values based on surrogates are only valid to the degree of the approximation quality of the surrogate model. Such falsified values can be avoided by adapting the calculation of the BME value, as proposed in \citet{mohammadi2018bayesian}. We will consider that the prediction of the surrogate model $\tilde{M}_k$ contains an approximation error $E_k$. We consider it to be independent of the measurement error $\epsilon$ (because $E_k$ and $\epsilon$ have no interaction), so that $M_k = \tilde{M}_k + E_k$. Therefore $p(\boldsymbol{y}_0 | \tilde{M}_k + E_k, \boldsymbol{\omega}) = p(\boldsymbol{y}_0 | \tilde{M}_k, \boldsymbol{\omega})~p(M_k | \tilde{M}_k, \boldsymbol{\omega})$ and the BME value in equation \eqref{eq:bme} can be rewritten as:

\begin{align}
    p(\boldsymbol{y}_0|M_k) = \int_\Omega p(\boldsymbol{y}_0 | \tilde{M}_k, \boldsymbol{\omega})~p(M_k | \tilde{M}_k, \boldsymbol{\omega})~p(\boldsymbol{\omega} | M_k)~d\boldsymbol{\omega} \text{,}
    \label{eq:helpBME}
\end{align}

where $p(M_k | \tilde{M}_k, \boldsymbol{\omega})$ is the likelihood function that indicates how well the original model prediction based on the model parameter realization $\boldsymbol{\omega}$ matches the corresponding surrogate model prediction:

\begin{align}
    p(M_k| \tilde{M}_k, \boldsymbol{\omega}) &= (2\pi)^{-N_\text{s}/2}|\boldsymbol{S}|^{-1/2} \nonumber
    \\
    &\cdot \exp \left( -\frac{1}{2} (\boldsymbol{y}_k(\boldsymbol{\omega}) - \tilde{\boldsymbol{y}}_k(\boldsymbol{\omega}))^T \boldsymbol{S}^{-1} (\boldsymbol{y}_k(\boldsymbol{\omega}) - \tilde{\boldsymbol{y}}_k(\boldsymbol{\omega})) \right) \text{,}
    \label{eq:newlikelihood}
\end{align}

with the predictions $\boldsymbol{y}_k$ of the original model and $\tilde{\boldsymbol{y}}_k$ of the surrogate model and the covariance matrix $\boldsymbol{S}$ of approximation errors.

Following the derivation in \citet{mohammadi2018bayesian}, we  obtain the corrected BME value for the original model, computed on the basis of the reduced model:

\begin{align}
    p(\boldsymbol{y}_0|M_k) = p(\boldsymbol{y}_0 | \tilde{M}_k) \cdot \int_\Omega p(M_k | \tilde{M}_k, \boldsymbol{\omega})~p(\boldsymbol{\omega} | \tilde{M}_k, \boldsymbol{y}_0)~d\boldsymbol{\omega} \text{.}
    \label{eq:newBME}
\end{align}

Equation \eqref{eq:newBME} shows clearly how the BME value of the original model ($\text{BME}_\text{OM}$) can be calculated from the BME value of the surrogate model ($\text{BME}_\text{SM}$):

\begin{align}
    \text{BME}_\text{OM} = \text{BME}_\text{SM} \cdot \text{Weight}_\text{SM} \text{,}
    \label{eq:bme_corrected}
\end{align}

with

\begin{align}
    \text{BME}_\text{OM} &= p(\boldsymbol{y}_0|M_k) \text{,} \nonumber
    \\
    \text{BME}_\text{SM} &= p(\boldsymbol{y}_0 | \tilde{M}_k)~\text{and} \nonumber
    \\
    \text{Weight}_\text{SM} &= \int_\Omega p(M_k | \tilde{M}_k, \boldsymbol{\omega})~p(\boldsymbol{\omega} | \tilde{M}_k, \boldsymbol{y}_0)~d\boldsymbol{\omega} \text{,}
    \label{eq:bme_corrected_variables}
\end{align}

where the $\text{BME}_\text{SM}$ value can be computed as described in the previous section, using the surrogate model $\tilde{M}_k$ instead of the original model $M_k$.

The correction factor Weight$_\text{SM}$ requires an integration over the whole parameter space $\Omega$ and its computation via Monte Carlo Integration is not feasible due to the high computational costs of the original model. Therefore, the correction factor can be estimated at those collocation points $\boldsymbol{\omega}^*$ that were used to construct the surrogate model:

\begin{align}
    \text{Weight}_\text{SM} \approx \sum\limits_{i=1}^P p(M_k | \tilde{M}_k, \boldsymbol{\omega}_i^*)~p(\boldsymbol{\omega}_i^* | \tilde{M}_k, \boldsymbol{y}_0)\text{,}
    \label{eq:weight}
\end{align}

where $P$ is the number of collocation points.

\subsection{Bayesian Model Justifiability Analysis}
\label{sec:methods_matrix}

In order to not only compare the models against the measurement data, \citet{schoniger2015finding} suggested a so-called model justifiability analysis, in which the competing models are tested against each other in a synthetic setup omitting the measurement data. The results of the justifiability analysis can help to decide whether the apparent by most appropriate model from the conventional BMS analysis is really the best model in the set or whether this model is only optimal given the limited amount of available measurement data \citep{schoniger2015finding}. Additionally, the justifiability analysis provides insights about similarities among the tested models. 

To perform the justifiability analysis, we will generate the so-called model confusion matrix \citep{schoniger2015finding} that is typically used in the field of statistical classification \citep[e.g.,][]{alpaydin_introduction_2004}. Confusion matrices compare the actual and the predicted classification, visualizing whether an object is misclassified (``confused''). In that way, we can recognize whether a model is able to distinguish its own predictions from the ones of its competitors. To do so, we calculate the Bayesian model weights for all models adopting equation \eqref{eq:bayes}.

However, instead of using the measurement data $\boldsymbol{y}_0$, each of the competing models generates a finite series of prior predictions that serve as realizations of the ``synthetic truth''. Thus, we generate $N_{\text{MC}}$ synthetic data sets of each model based on samples of its prior parameter distributions. Then, each synthetic data set is compared to the competing models by first computing the likelihood function as described in equation \eqref{eq:likelihood}, for example of the single realization $i$ of model $M_k$ based on the data set $j$ of model $M_l$. The BME value can be obtained by calculating the mean of all likelihoods $p(M_{l,j}|M_k)$ of model $M_k$ given this single realization $j$ of model $M_l$. The resulting model confusion matrix has the size $N_\text{m} \times N_\text{m}$, for $N_\text{m}$ competing models.

To execute both steps of model testing ((1) BMS testing against measurements and (2) justifiability analysis testing models against each other) simultaneously, we add the measurement data to our model set, i.e. we add it as a new row and column to the confusion matrix.

\begin{figure}[hptb]
    \centering
    \includegraphics[width = 0.65\textwidth]{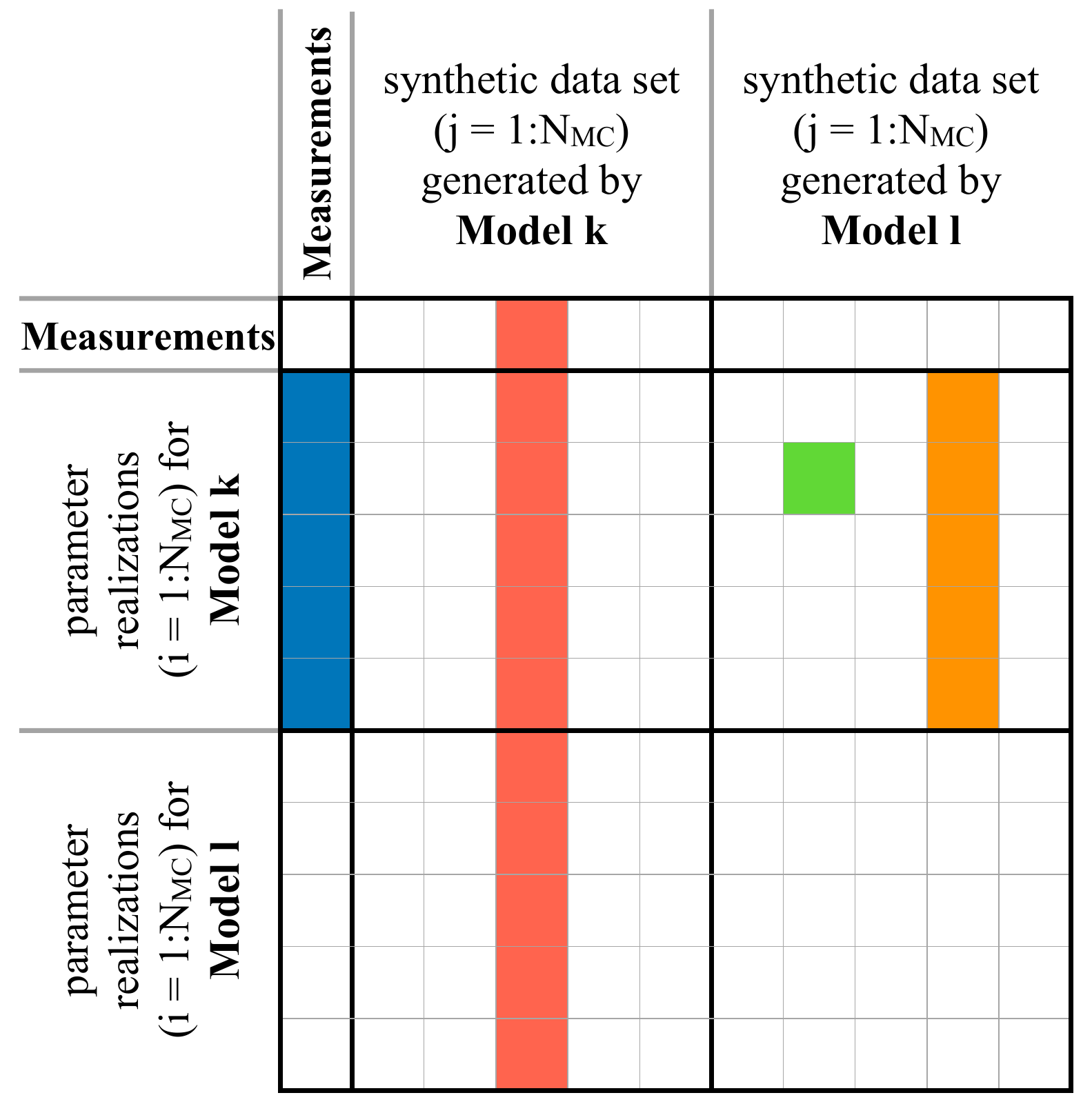}
    \captionof{figure}{Schematic illustration of constructing the model confusion matrix.}
    \label{fig:bme_matrix}
\end{figure}

 A schematic illustration of its construction is given in Figure \ref{fig:bme_matrix}, whereby the model confusion matrix is extended by the standard BMS procedure.
 
The blue box in Figure \ref{fig:bme_matrix} represents a standard BMS procedure where the model $M_k$ has been tested against the measurement data. This entry can be obtained from \eqref{eq:bme}, using Monte Carlo Integration for $p(\boldsymbol{y}_0 | M_k)$ as in \eqref{eq:MC}.
The green box in Figure \ref{fig:bme_matrix} reflects the likelihood of a single realization of model $M_k$ given a single realization of the reference model $M_l$, which currently serves to stand as synthetic truth. The orange box in Figure \ref{fig:bme_matrix} shows the average likelihood (BME) of model $M_k$ given a single realization of the reference model $M_l$. This BME value is normalized by the sum of the BME values of all models given a single realization of the synthetic truth (red box), yielding a posterior model weight $p(M_l|M_{k,j})$ with the reference model $M_k$. The bold boxes in Figure \ref{fig:bme_matrix} illustrate these averaged posterior weights over all synthetic data sets of the reference model $M_k$. The bold boxes of one column contain the expected posterior weights ($PW$) of all models given that model $M_k$ is true. One entry can be computed as follows:

\begin{align}
PW_{lk} &=  \frac{1}{N_{\text{MC}}} \sum\limits_{j=1}^{N_{\text{MC}}} p(M_l|M_{k,j})
\\      &=  \frac{1}{N_{\text{MC}}^2} \sum\limits_{j=1}^{N_{\text{MC}}} \sum\limits_{i=1}^{N_{\text{MC}}} p(M_{l,i}|M_{k,j})\text{,}
\label{eq:bme_entry}
\end{align}

whereby the averaged BME value $\left(\sum\limits_{i=1}^{N_{\text{MC}}} p(M_{l,i}|M_{k,j})\right)$ in eq. \eqref{eq:bme_entry} is not normalized for the sake of readability.

The resulting extended model confusion matrix consists only of these entries, i.e. the bold boxes and therefore has the size $(N_\text{m}+1) \times (N_\text{m}+1)$, for $N_\text{m}$ competing models and the measurement data.

The main diagonal entries reflect how good each model identifies itself as the data-generating process, given a certain data set size. The values of the diagonal entries should be equal to $1.00$ with an infinite data set size. However, for finite data sets, models might ``confuse'' their own predictions (misclassification) with the ones of competing models due the two following reasons. (1) Two models are actually highly similar. (2) One model has a high goodness-of-fit to the reference data, but also a high variability in its predictions. The BMS framework punishes this high variability with a lower model weight. Thus, a scenario of a more and a less variable model, which fits the reference data worse than the more variable one, might lead to similar model weights. When more synthetic data is used, the more variable model will receive a higher weight, as its variability becomes more justifiable, while the weight of the less variable model will decrease \citep[][]{hoge2018primer, hoge2019hydrologist}.

The off-diagonal entries of the model confusion matrix reflect the similarity between pairs of models. This can be useful when comparing possible simplifications to a detailed reference model \citep{schaferstrategies}. With the aid of the model confusion matrix, it is possible to identify the model that yields results that are most similar to the reference model, at reduced computational costs.

\subsection{aPC-Based Bayesian Model Justifiability Analysis}
\label{sec:methods_bayesian_pce_models}

We will combine the methodologies from Sections \ref{sec:methods_bayesian_pce} and \ref{sec:methods_matrix} towards an aPC-based Bayesian model justifiability analysis, where models are mutually tested against each other.
To do so, we will consider two models, model $M_k$ and model $M_l$. The comparison of two models implies that one model, $M_l$ in this case, is assumed to be the data-generating process. Instead of computing the BME value for the original models $p(M_l|M_k)$, we have to calculate the BME value $p(\tilde{M}_l|\tilde{M}_k)$ of the surrogate models. Similar to Section \ref{sec:methods_bayesian_pce}, we assume that each surrogate representation of each analyzed model contains an approximation error: $M_k = \tilde{M}_k + E_k$ and $M_l = \tilde{M}_l + E_l$. Therefore, equation \eqref{eq:newBME} can be rewritten as:

\begin{align}
    p(M_l|M_k) = p(M_l | \tilde{M}_k) \cdot \int_\Omega p(M_k | \tilde{M}_k, \boldsymbol{\omega})~p(\boldsymbol{\omega} | \tilde{M}_k, M_l)~d\boldsymbol{\omega} \text{.}
    \label{eq:helpBME_model_part1}
\end{align}

In the next step, we focus on the term $p(M_l | \tilde{M}_k)$, considering  $M_l = \tilde{M}_l + E_l$ leads us to

\begin{align}
    p(M_l | \tilde{M}_k) = \int_\Omega p(\tilde{M}_l | \tilde{M}_k, \boldsymbol{\omega})~p(M_l | \tilde{M}_l, \boldsymbol{\omega}_k)~p(\boldsymbol{\omega} | \tilde{M}_k)~d\boldsymbol{\omega} \text{.}
    \label{eq:helpBME_model_part2}
\end{align}

Multiplying and dividing the right-hand side of \eqref{eq:helpBME_model_part2} with $p(\tilde{M}_l | \tilde{M}_k)$ and applying Bayes' theorem yields

\begin{align}
    p(M_l | \tilde{M}_k) = p(\tilde{M}_l | \tilde{M}_k) \cdot \int_\Omega p(M_l | \tilde{M}_l, \boldsymbol{\omega})~p(\boldsymbol{\omega} | \tilde{M}_k, \tilde{M}_l)~d\boldsymbol{\omega} \text{.}
    \label{eq:newBME_model_part2}
\end{align}

When inserting \eqref{eq:newBME_model_part2} into \eqref{eq:helpBME_model_part1}, we obtain

\begin{align}
    p(M_l|M_k) = p(\tilde{M}_l | \tilde{M}_k) \cdot &\int_\Omega p(M_l | \tilde{M}_l, \boldsymbol{\omega})~p(\boldsymbol{\omega} | \tilde{M}_k, \tilde{M}_l)~d\boldsymbol{\omega} \nonumber
    \\
    \cdot &\int_\Omega p(M_k | \tilde{M}_k, \boldsymbol{\omega})~p(\boldsymbol{\omega} | \tilde{M}_k, M_l)~d\boldsymbol{\omega},
    \label{eq:newBME_model}
\end{align}

or

\begin{align}
    \text{BME}_\text{OMOM} = \text{BME}_\text{SMSM} \cdot \text{Weight}_\text{SM1} \cdot \text{Weight}_\text{SM2}\text{,}
    \label{eq:bme_model_corrected}
\end{align}

with

\begin{align}
    \text{BME}_\text{OMOM} &= p(M_l|M_k) \nonumber
    \\
    \text{BME}_\text{SMSM} &= p(\tilde{M}_l | \tilde{M}_k) \nonumber
    \\
    \text{Weight}_\text{SM1} &= \int_\Omega p(M_l | \tilde{M}_l, \boldsymbol{\omega})~p(\boldsymbol{\omega} | \tilde{M}_k, \tilde{M}_l)~d\boldsymbol{\omega} \nonumber
    \\
    \text{Weight}_\text{SM2} &= \int_\Omega p(M_k | \tilde{M}_k, \boldsymbol{\omega})~p(\boldsymbol{\omega} | \tilde{M}_k, M_l)~d\boldsymbol{\omega} \text{,}
    \label{eq:bme_model_corrected_variables}
\end{align}

whereby $\text{BME}_\text{OMOM}$ corresponds to the BME value when comparing two original models and $\text{BME}_\text{SMSM}$ to the BME value when comparing two surrogate models.
The value of $\text{BME}_\text{SMSM}$ can be computed in the same way as proposed in \eqref{eq:bme} via Monte Carlo integration in \eqref{eq:MC} with the likelihood function defined in \eqref{eq:likelihood}, using the prediction of model $M_l$ evaluated on a certain model parameter vector $\boldsymbol{\omega}$ instead of the measurement data $\boldsymbol{y}_0$. The collocation points $\boldsymbol{\omega}^*$ can be employed again similarly to Section \ref{sec:methods_bayesian_pce} to compute the correction factors for both models:

\begin{align}
    \text{Weight}_\text{SM1} &\approx \sum\limits_{i=1}^P p(M_l | \tilde{M}_l, \boldsymbol{\omega}_i^*)~p(\boldsymbol{\omega}_i^* | \tilde{M}_k, \tilde{M}_l) \nonumber
    \\
    \text{Weight}_\text{SM2} &\approx \sum\limits_{i=1}^P p(M_k | \tilde{M}_k, \boldsymbol{\omega}_i^*)~p(\boldsymbol{\omega}_i^* | \tilde{M}_k, M_l) \text{.}
    \label{eq:weights_model} 
\end{align}

Moreover, since the model confusion matrix in the Bayesian model justifiability framework compares the original models as well, we have to account for the approximation of these models with the surrogates. As the weights $\text{Weight}_\text{SM1}$ and $\text{Weight}_\text{SM2}$ are not dependent on a single parameter realization, the overall posterior weights of the model confusion matrix can be corrected in the same way as the BME values. To this end, the posterior values ($PW$) of the model confusion matrix from equation \eqref{eq:bme_entry} need to be multiplied with the two correction factors $\text{Weight}_\text{SM1}$ and $\text{Weight}_\text{SM2}$ from \eqref{eq:weights_model}: 

\begin{align}
  PW_{lk} &=  \frac{1}{N_{\text{MC}}} \sum\limits_{j=1}^{N_{\text{MC}}} p(M_l|M_{k,j}) \nonumber
  \\
  &=   \frac{1}{N_{\text{MC}}} \sum\limits_{j=1}^{N_{\text{MC}}} p(\tilde{M}_l|\tilde{M}_{k,j}) \cdot \text{Weight}_\text{SM1} \cdot \text{Weight}_\text{SM2} \text{,}
  \label{eq:BME_entry}
\end{align}

 where SM1 $= \tilde{M}_l$ and SM2 $= \tilde{M}_k$.

\section{Biogeochemical Processes in Porous Media}
\label{sec:models}

\subsection{Microbially Induced Calcite Precipitation}
\label{sec:models_micp}
Microbially induced calcite precipitation (MICP) is a typical biogeochemical process.
When conceptualizing MICP in porous media, various phases are involved: there are at least three solid phases (biofilm, calcite and unreactive solid material), water and possibly another fluid phase, e.g. gas. Additionally, at least calcium, inorganic carbon, and urea are considered as dissolved components in the water phase, the complete list of components can be found in \citet{hommel2015revised}.

MICP is a reactive transport process consisting of three main parts: (1) adhesion of biomass on surfaces, detachment of the biomass from the biofilm as well as growth and decay of the biomass, (2) urea hydrolysis that alters the geochemistry and (3) precipitation and dissolution of calcite.

\begin{figure}[hptb]
    \centering
    \includegraphics[width = 0.8\textwidth]{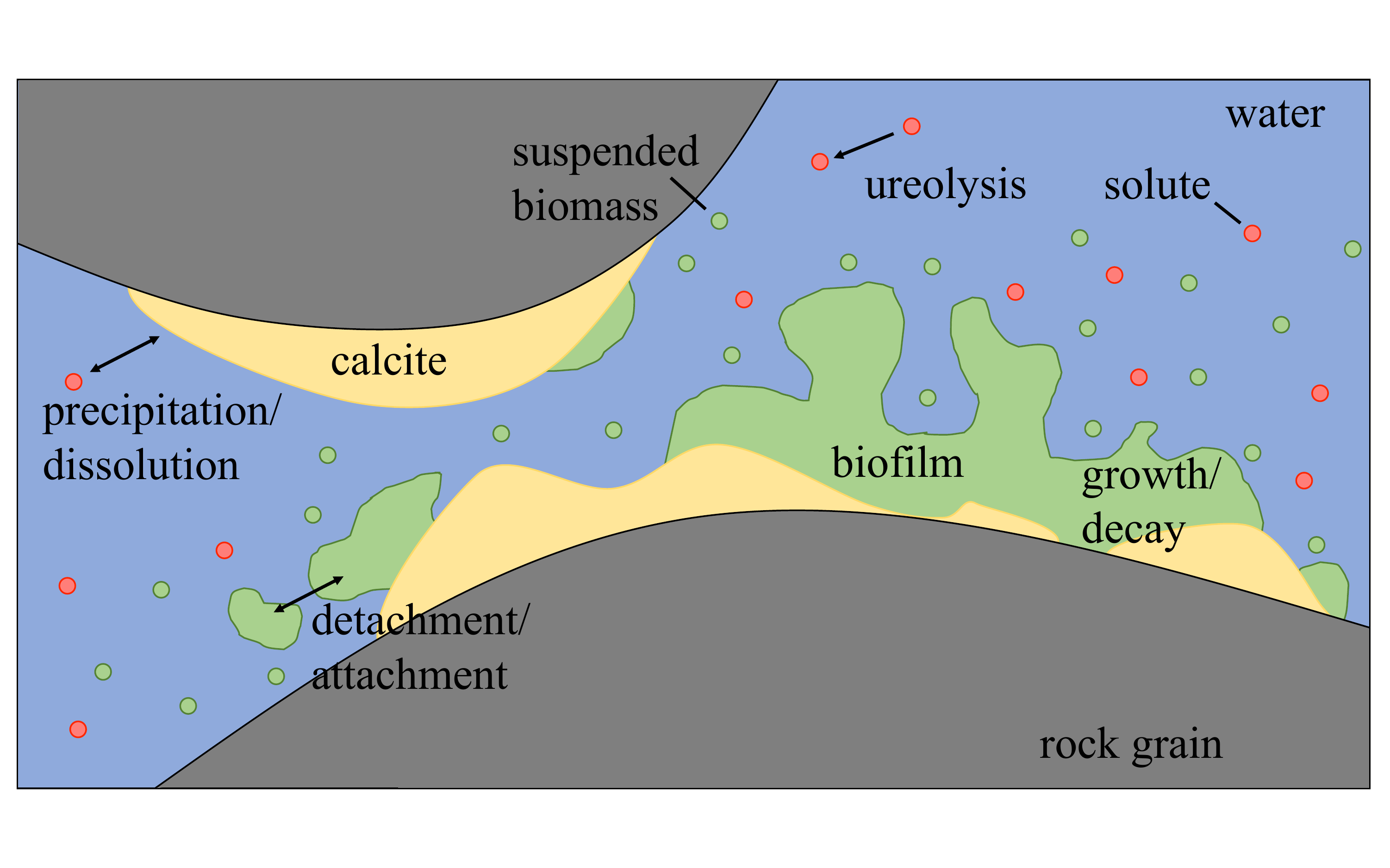}
    \captionof{figure}{Schematic view of relevant processes and phases during MICP after \citet{hommel2015revised}.}
    \label{fig:micp_processes}
\end{figure}

A visualization of the MICP process is shown in Figure \ref{fig:micp_processes}.

S. pasteurii are bacteria that are able to produce the enzyme urease and to decompose urea into carbonic acid and ammonia with the aid of urease.

In aqueous solution, the ammonia reacts with the contained H$^+$ ions.
As a result, the pH value increases so that the carbonic acid decomposes into H$^+$ ions and carbonate ions, while the concentration of dissolved carbonate increases.

If calcium ions are provided, it comes to a reaction with the carbonate ions and calcite precipitates.

Shortly, all together this leads to the following MICP reaction equation \citep{hommel2015revised}:
\begin{reaction}
\ce{CO(NH2)2 + 2H2O + Ca^2+ ->[Urease] 2NH4^+ + CaCO3} $\downarrow$ \text{.}
\label{eq:micp}
\end{reaction}

\subsection{Experimental Setup}
\label{sec:models_experiment}
\begin{figure}[hptb]
    \centering
    \includegraphics[width = 1\textwidth]{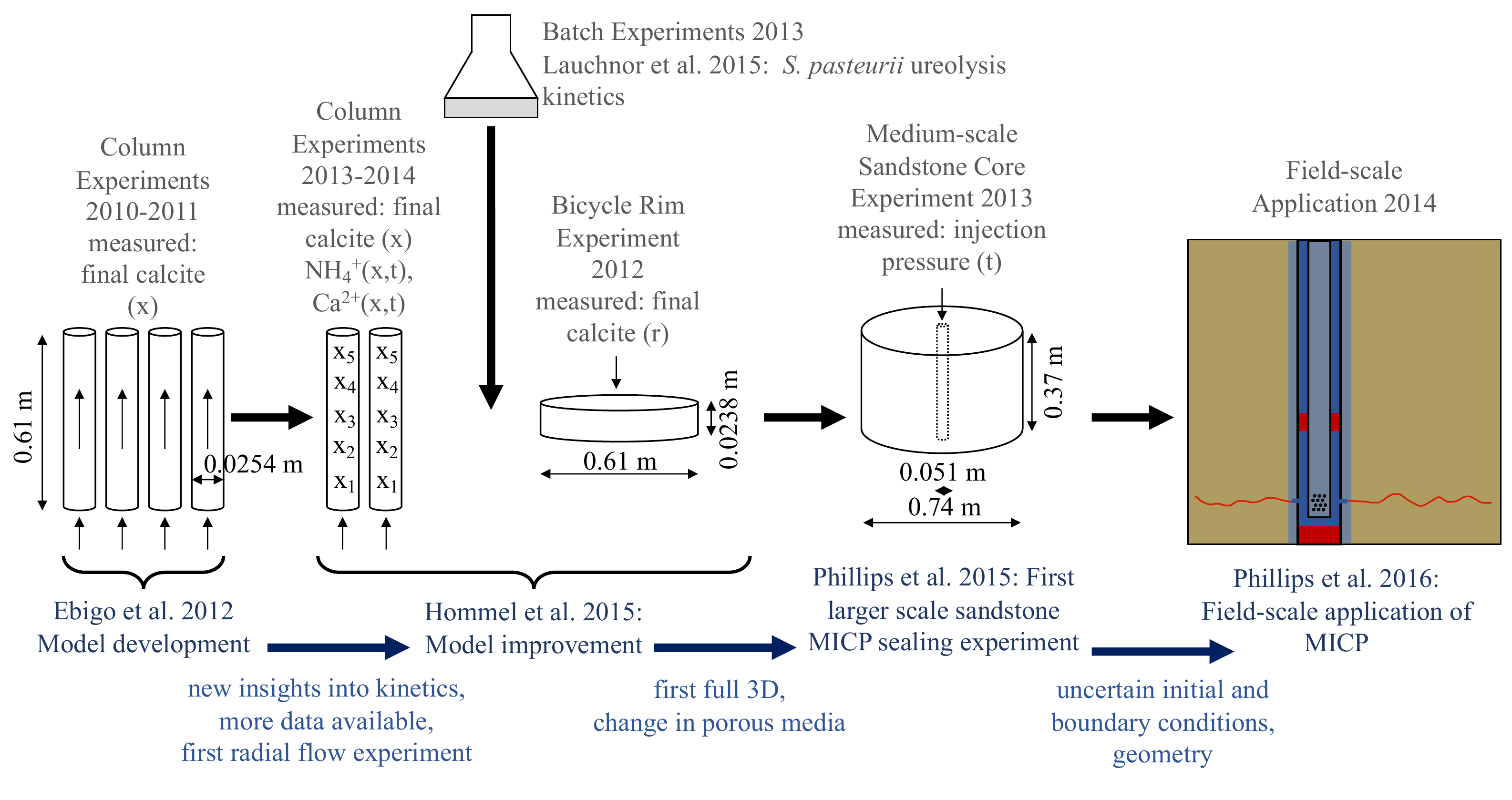}
    \captionof{figure}{Model and experiment development involved in preparation for the field-scale application after \citet{cunningham2019field}.}
    \label{fig:micp_experiment}
\end{figure}

The analyzed MICP experiment is described in detail in \citet{hommel2015revised} (there, see experiment ``D1''). It describes a sand-filled column that is \SI{61}{cm} high with a diameter of \SI{2.54}{cm}. In the beginning of the experiment, bacteria are injected at the bottom of the column, until a sufficient amount is accumulated and a biofilm is established. Then, the biofilm is fed once again (bacteria are injected again). From now on, calcium and urea are injected repeatedly every \SI{24}{hours}. This allows the mineralization reactions to take place. That period is followed by another injection of biomass to revive the microorganisms (or rather feed them once again) \citep{hommel2015revised}, before the next injections of calcium and urea start over. A full model and experiment development after  \citet{cunningham2019field} is shown in Figure \ref{fig:micp_experiment}.

The models predict the calcium and calcite over space and time (\ref{sec:models_set}). The predictions are compared to measurement data as well as among each other. In order to receive comparable results, only spatial and temporal points where measurement data is available as well are used when comparing models among each other. These data points differ for calcium and calcite. For the calcite content, there is only measurement data available at the end of the experiment, which is after \SI{3203460}{seconds} (about \SI{890}{hours} or \SI{37}{days}). The calcium concentration is measured at 35 different data points in time.
Therefore, calcium is injected at 6 ``main points'' in time, the so-called pulses, namely after 151.35, 218.85, 290.85, 626.85, 698.85 and \SI{866.85}{hours}. At these points, the concentration is measured and additionally respectively after half an hour, after one, two, three and four hours, except for pulse 22, where no measurement is available after \SI{3}{hours}, which results in 35 temporal points. The exact times of measurement after the first injection can be taken from Table \ref{table:temp_points}.
\begin{table}[H]
    \centering
    \begin{tabular}{rcccccc}
        \hline\noalign{\smallskip}
        \diagbox{after pulse}{pulse number} & 5 & 7 & 10 & 22 & 24 & 30
        \\
        \noalign{\smallskip}\hline\noalign{\smallskip}
        \SI{0}{hours} & 151.35 & 218.85 & 290.85 & 626.85 & 698.85 & 866.85
        \\
        \SI{0.5}{hours} & 151.85 & 219.35 & 291.35 & 627.35 & 699.35 & 867.35
        \\
        \SI{1}{hour} & 152.35 & 219.85 & 291.85 & 627.85 & 699.85 & 867.85
        \\
        \SI{2}{hours} & 153.35 & 220.85 & 292.85 & 628.85 & 700.85 & 868.85
        \\
        \SI{3}{hours} & 154.35 & 221.85 & 293.85 & - & 701.85 & 869.85
        \\
        \SI{4}{hours} & 155.35 & 222.85 & 294.85 & 630.85 & 702.85 & 870.85
        \\
        \noalign{\smallskip}\hline
    \end{tabular}
    \caption{Times in hours for measurement of the calcium concentration}
    \label{table:temp_points}
\end{table}
There are eight measurement locations for the calcite concentration, located at 3.81, 11.43, 19.05, 26.67, 34.29, 41.91, 49.53 and \SI{57.15}{cm} distance from the bottom. For the calcium concentration, there are only five spatial measurement points located at 10.16, 20.32, 30.48, 39.37 and \SI{49.53}{cm} distance from the bottom. The measurement locations in the models are evenly distributed at a respective distance of half an inch (\SI{1.27}{cm}).

\subsection{Conceptual Models and Related Uncertainty}
\label{sec:models_set}

We analyze three models for MICP that describe biogeochemical processes in porous media provided by \citet{hommel2015revised, hommel2016finding}. For detailed explanation of their equations and the used numerical schemes, we refer to that original publication.

An \textless Intel(R) Xeon(R) CPU E5-2680 v2 @2.80 GHz, 40 Cores\textgreater~machine was used for the model evaluations. The computational effort for the most detailed MICP model, referred to as \textit{full complexity} model, is extremely high with a run time between \SI{16} and \SI{42}{hours}, depending on the respective model parameter set. The exact costs are dependent on the model parameter set chosen for the evaluation, since the time stepping varies adaptively. Therefore, \citet{hommel2015revised} suggest two simplifications of the \textit{full complexity} model $M_\text{FC}$ using the certain physical assumptions.
\begin{itemize}
    \item \textit{initial biofilm} model ($M_\text{IB}$): the suspended biomass is ignored and the biofilm is already established at the beginning of the experiment.
    \item  \textit{simple chemistry} model ($M_\text{SC}$): all the urea injected to the system is assumed to precipitate as calcite. The precipitation occurs immediately as described in the overall reaction equation \rxnref{eq:micp} \citep{hommel2016finding}.
\end{itemize}
As described in Section \ref{sec:models_experiment}, the experiment starts with a biomass injection and a waiting period until the biofilm is established. The \textit{initial biofilm} model $M_\text{IB}$ omits this part of the simulation under the assumption that the biofilm is already established in the beginning of the experiment and the attachment periods of the biomass are not simulated \citep{hommel2016finding}. The \textit{simple chemistry} model $M_\text{SC}$ simplifies the reactions of urea. The model makes the assumption, that all the urea put into the system completely reacts to calcite. Therefore, there is no need for computing the ureolysis rate and the precipitation rate, or either the expensive-to-calculate saturation state and carbonate and calcium activities \citep{hommel2016finding}. The computational time of the \textit{initial biofilm} model $M_\text{IB}$ still remains high and and is only slightly lower than for the \textit{full complexity} model on the same computational cluster. The strong assumptions in the \textit{simple chemistry} model $M_\text{SC}$ allow to obtain results of one model run after 40 minutes using the same computational cluster. 

Apart from decreasing the computational cost, model simplification reduces parametric uncertainty. A too detailed (too complex) model with many parameters and without enough calibration data (and therefore parametric uncertainty) results in a high predictive variance (i.e. uncertainty) of the model.

Models should generally be ``as simple as possible, as complex as necessary'' (principle of parsimony) \citep{hoge2018primer} to prevent overfitting \citep[e.g.][]{babu_resampling_2011, lever_model_2016}.
The considered parameters in the following were previously identified as the most uncertain parameters of the MICP models in \citet{hommel2015revised}:
\begin{itemize}
    \item the coefficient for preferential attachment to biomass $c_\text{a,1}$, $\left[\si{\per\second}\right]$
    \item the coefficient for attachment to arbitrary surfaces $c_\text{a,2}$, $\left[\si{\per\second}\right]$
    \item the mass density of biofilm $\rho_\text{f}$, $\left[\si{kg\per\cubic\meter}\right]$
    \item the enzyme content of biomass $k_\text{ub}$, $\left[\si{kg\per kg}\right]$.
\end{itemize}
As the \textit{initial biofilm} model $M_\text{IB}$ assumes that there are no attachment periods, it is only dependent on the model parameters $\rho_\text{f}$ and $k_\text{ub}$. The \textit{full complexity} model $M_\text{FC}$ and \textit{simple chemistry} model $M_\text{SC}$ are both dependent on all four model parameters. Following the physically possible range of the considered uncertain parameters, we assume that all of the model parameters are uniformly distributed in the intervals shown in Table \ref{table:parameter_intervals}.
\begin{table}[H]
    \centering
    \begin{tabular}{rl}
        \hline\noalign{\smallskip}
        model parameter & interval
        \\
        \noalign{\smallskip}\hline\noalign{\smallskip}
        $c_\text{a,1}$ & $\left[\SI{1e-10}{\per\second}, \SI{1e-7}{\per\second} \right]$ 
        \\
        $c_\text{a,2}$ & $\left[\SI{1e-10}{\per\second}, \SI{1e-6}{\per\second} \right]$ 
        \\
        $\rho_\text{f}$ & $\left[\SI{1}{kg\per\cubic\meter}, \SI{15}{kg\per\cubic\meter} \right]$ 
        \\
        $k_\text{ub}$ & $\left[\SI{1e-5}{kg\per kg}, \SI{5e-4}{kg\per kg} \right]$ 
        \\
        \noalign{\smallskip}\hline
    \end{tabular}
    \caption{Intervals for the model parameters}
    \label{table:parameter_intervals}
\end{table}

\subsection{Implementation Details of the Surrogate Models}
\label{sec:models_impl}

We construct three surrogate models for the three competing MICP models described in Section \ref{sec:models_set} using a $d = 2$ order aPC expansion according to the prior distributions presented in Table \ref{table:parameter_intervals}. For this purpose, the three original models will be evaluated $\displaystyle D = (N_p+d)!/(N_p!d!)-1$ times according to Section \ref{sec:methods_pce}. Since the $D$ evaluations for the construction of the surrogate models are independent, these model runs were parallelized. Further, we refine each of the three surrogates using iterative Bayesian updating of the aPC representation according to Section \ref{sec:methods_iterative}. Here, we restrict the number of Bayesian updates to ten due to the high computational demand and previous experience (see e.g. \citep{beckers2020bayesian}), so that $\displaystyle P_\text{end} = D+10 = (N_p+d)!/(N_p!d!)-1 + +10$. During the Bayesian updating, we consider the standard deviation of measurement errors $\epsilon$ at each point in space (and time) equal to 20\% of the associated measurement value for both the calcite content and the calcium concentration.

\section{Justifiability Analysis of Biogeochemical Models in Porous Media}
\label{sec:results}

\subsection{aPC-Based Representation of MICP Models}
\label{sec:results_approx}

\begin{figure}[hptb]
    \centering
    \includegraphics[width = 1\textwidth]{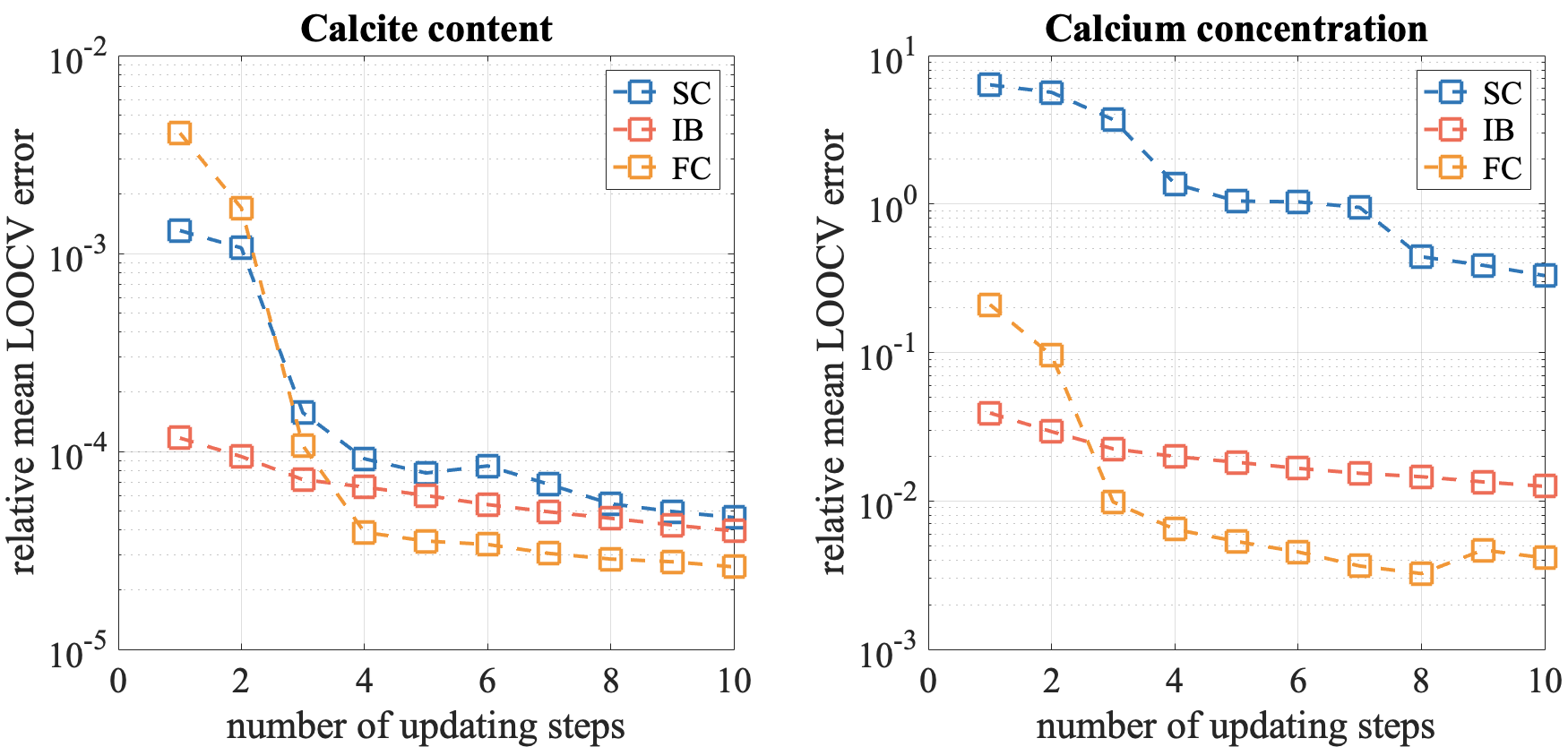}
    \captionof{figure}{Relative mean LOOCV errors for calcite content and calcium concentration with increasing number of updates.}
    \label{fig:rel_loocv_evolution}
\end{figure}

Equation (\ref{eq:MittelLOOCV}) provides errors of the surrogate models for every point in space and time due to the structure of equation (\ref{eq:modelapprox}). As every point in space and time has its own surrogate model, there are $5 \cdot 35 \cdot 10 = 1750$ LOOCV errors (5 spatial and 35 temporal points, 10 updating steps) computed for calcium and $8 \cdot 10$ for calcite (8 spatial points, 10 updating steps) in the analyzed set up. The LOOCV error is computed after the primal construction of the surrogate models and during the iterative Bayesian updating. In order to visualize the errors, we will average the respective values over space (and time) after every updating step. In order to compare the LOOCV error of the surrogate models for calcium and calcite, the relative errors must be considered, since the two quantities of interest (Calcite content [\%] and calcium concentration [$\si{mol\per\cubic\meter}$]) are in different orders of magnitude. For this purpose, they were normalized to the mean output value, as shown in Figure \ref{fig:rel_loocv_evolution}.

The relative mean LOOCV errors before the first update are not considered in this figure to get a better visualization, since this error is significantly higher than the ones after the updates. First of all, the figure shows that the surrogate error for calcite decreases more strongly than the error for calcium. It is also remarkable that the error for all models for calcite is in a similar order of magnitude. This means that all surrogate models are of a comparable quality for the calcite content. For calcium, the error of the \textit{simple chemistry} model $M_\text{SC}$ is significantly larger than the one for the other two surrogate models. This can occur if one uses Bayesian updating and wants to improve the models only in the region of the measurement data. This means the surrogate model is similar to the original one in the region of the measurement data, but it deviates a lot from the original model in other regions (not part of the measurement points). This results in a higher overall LOOCV error. The larger error of the surrogate model is compensated later by the newly introduced correction factor in Section  \ref{sec:methods_bayesian_pce}.

Furthermore, the errors for calcite are in a range of $[\num{2e-5}, \num{6e-5}]$ after the last update and those for calcium are in a range of $[\num{4e-3}, \num{4e-1}]$. Accordingly, the worst surrogate response for calcite is still better than the best one for calcium. This indicates that the surrogate models for the calcite content as a whole are better with respect to the LOOCV error than those for the calcium concentration.

\subsection{aPC-Based Justifiability Analysis for MICP Models}
\label{sec:results_comparison}

We will perform the aPC-based Bayesian model selection incorporating the measurement data and aPC-based Bayesian model justifiability analysis according to Sections \ref{sec:methods_bayesian_pce} and \ref{sec:methods_bayesian_pce_models} using the obtained surrogate representations of the three analyzed MICP models from Section \ref{sec:results_approx}. Following the justifiability analysis, we compute the model weights as stated in Section \ref{sec:methods_matrix} and adjust them with the novel correction factors from Sections \ref{sec:methods_bayesian_pce} and \ref{sec:methods_bayesian_pce_models} in a second stage. In order to justify the underlying physical assumptions behind the MICP models, we will assess the impact of the data set size onto BME values appearing in the Bayesian justifiability analysis. To do so, we start with only one spatial data point, then we use half of the available data set size and finally we include all of the spatial data points for calcium and calcite. This results in the following data set sizes $N_{D, \text{spatial}} \in \{1, 3, 5\}$ for calcium and $N_{D, \text{spatial}} \in \{1, 4, 8\}$ for calcite.

\subsubsection{aPCE-Based BMS and Justifiability Analysis}
\label{sec:results_bms}

\begin{figure}[hptb]
    \centering
    \includegraphics[width = 1\textwidth]{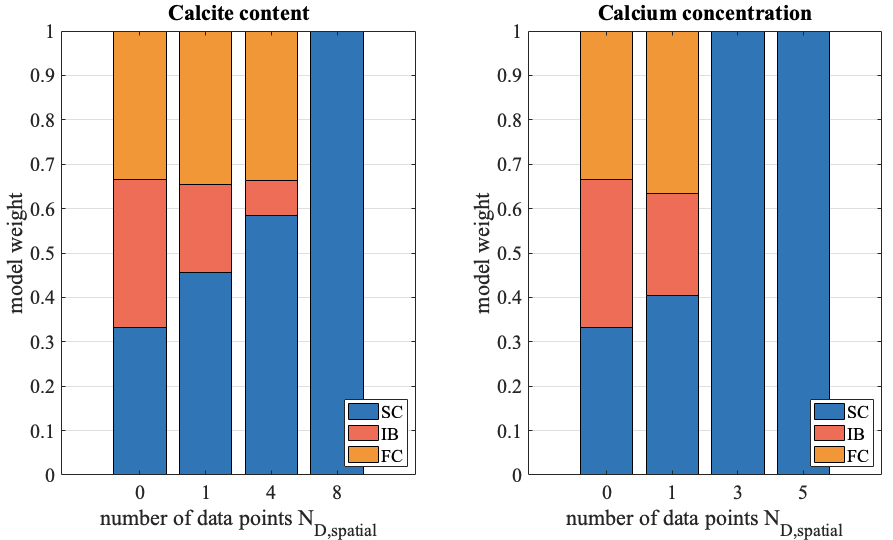}
    \captionof{figure}{Model weights for the prediction of calcite content and calcium concentration over increasing amount of used spatial data points $N_{D, \text{spatial}}$.}
    \label{fig:weight_bar}
\end{figure}

In a first stage, the conventional BMS analysis for measurement data is performed with results illustrated in Figure \ref{fig:weight_bar}.

One can observe that the \textit{simple chemistry} model $M_\text{SC}$ obtains the highest model weight (normalized BME value) for all data set sizes.
A model wins the competition either because of its low complexity or because of its goodness-of-fit to the measurement data (or both) \citep{schoniger2015finding}. These two aspects will be further investigated in a second stage, the justifiability analysis.

\begin{figure}[hptb]
    \centering
    \includegraphics[width = 1\textwidth]{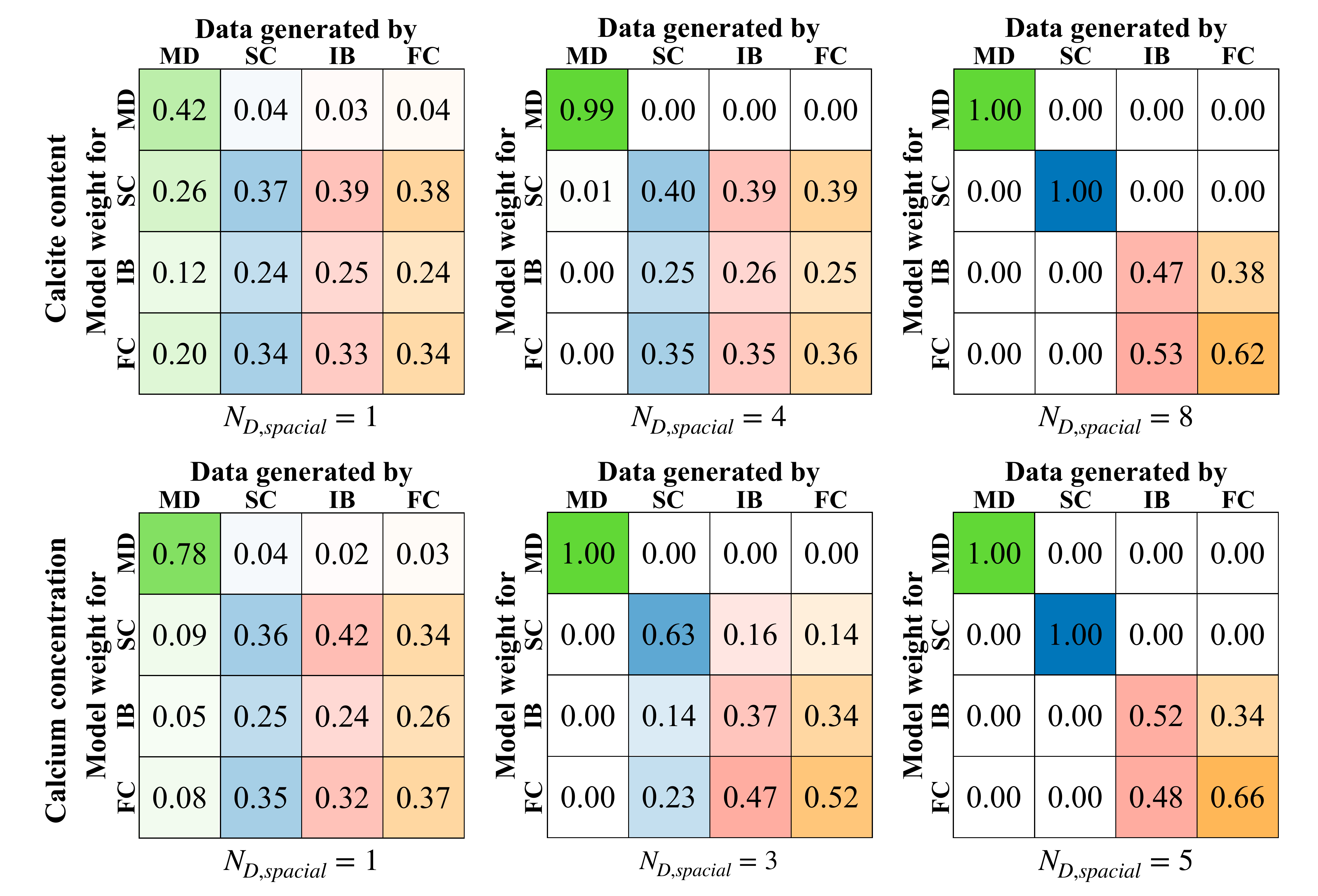}
    \captionof{figure}{Model confusion matrices for calcite content $[\%]$ and calcium concentration $[\si{mol\per\cubic\meter}]$ of the the three models and the measurement data (MD) over increasing amount of used spatial data points $N_{D, \text{spatial}}$.}
    \label{fig:post_matrices}
\end{figure}

Figure \ref{fig:post_matrices} shows the corresponding model confusion matrices for both the calcium concentration and the calcite content predictions. Each entry corresponds to the weight of one model, which is the probability that model $M_k$ (rows) is the data-generating process of the predictions made by model $M_l$ (columns) according to Bayes' theorem.

The main-diagonal entries of the model confusion matrix in  Figure \ref{fig:post_matrices} represent the models' ability to identify their own predictions. The higher the value of the main diagonal entry in Figure \ref{fig:post_matrices}, the higher probability of the model to identify itself as the data-generating process. The diagonal values increase when a bigger data set size is used, agreeing well with the Bayesian justifiability analysis discussed in \citep{schoniger2015finding}. The diagonal weight of the simplest model, the \textit{simple chemistry} model $M_\text{SC}$, is always the highest, independent of the data set size, which shows that the analysis identifies this model as data-generating, even if the data set is large and the model makes strong assumptions. For both the calcium and the calcite, the diagonal entries achieve the ``absolute majority'' of more than $0.50$ in favor of justifiability (except for the \textit{initial biofilm} model $M_\text{IB}$ for calcite) when taking the full data set into account. This means that the data set size is sufficient to justify the modeling concepts behind the considered models.

But even for the full data set, the \textit{full complexity} model $M_\text{FC}$ obtains a high weight when the \textit{initial biofilm} model $M_\text{IB}$ generates the data and vice versa. It follows that the \textit{initial biofilm} model $M_\text{IB}$ and the \textit{full complexity} model $M_\text{FC}$ confuse their predictions and are not confident in identifying their own predictions (the \textit{initial biofilm} model $M_\text{IB}$ for calcite is not even able to identify itself). However, only for the \textit{simple chemistry} model $M_\text{SC}$ the weight is $1.00$ and therefore its simplicity is perfectly supported with the full data set. The measurement data (MD) obtains a model weight of $1.00$ for the full data set too, since it is clearly able to identify itself with the full data set. The weights for the models with the measurement data as the data-generating process are strikingly low. In statistical terms, this means that all models are clearly rejected by the full data set. This fits with the conclusions drawn in \citet{hommel2015revised}, that there is at least one relevant process not yet implemented in ``sufficient detail'', which is necessary for better results.

\subsubsection{How Much Data Do We Need?}
\label{sec:results_data}

The matrices on the left in Figure \ref{fig:post_matrices} show that considering only one spatial data point is not sufficient, since the diagonal entries for calcite and calcium are all less than $0.50$ except for the measurement data for the calcium concentration. This means that there is no ``absolute majority'' in favor of justifiability for any model and even the measurement data of the calcite content is not able to identify itself (which is obvious since there is clearly a variance between the measurements at different spatial data points). The matrices also show that the simplest model (SC) obtains the highest weight of all three models when the data set size is small (principle of parsimony).

When using half of the data set, the simplest model $M_\text{SC}$ and the most complex model $M_\text{FC}$ for calcium receive an absolute majority with model weights of $0.63$ and $0.52$, while the data set size does not suffice for self-identification of the \textit{initial biofilm} model $M_\text{IB}$. The weight of $M_\text{IB}$ on the diagonal entry increases with an increasing data set size, but it never gains a weight greater than $0.5$. In contrast, the weight for $M_\text{IB}$ for the calcium concentration reaches the absolute majority, which means that the data set size is sufficient for self-identification and the physical model assumptions leading to simplifications are justifiable.

\begin{figure}[hptb]
    \centering
    \includegraphics[width = 1\textwidth]{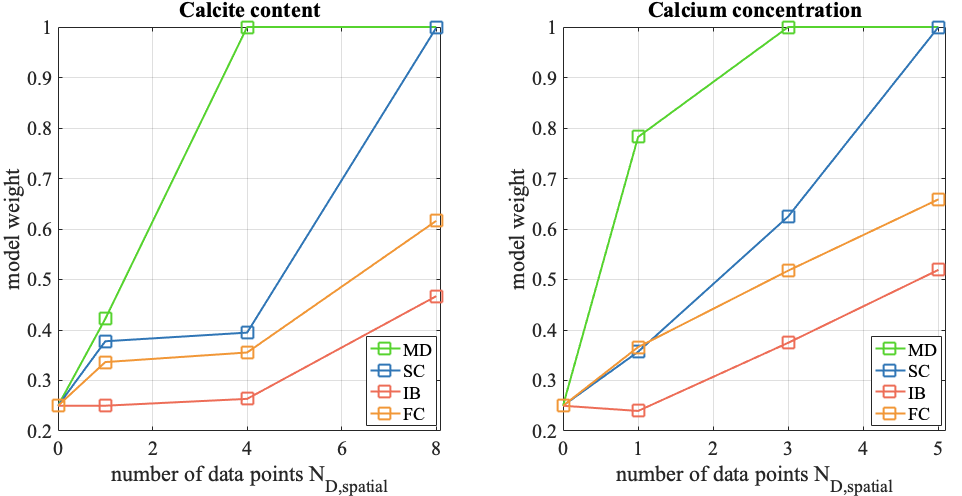}
    \captionof{figure}{Average model weights for the data-generating process of the two quantities of interest (calcite content and calcium concentration) of the the three models and the measurement data (MD) over increasing amount of used spatial data points $N_{D, \text{spatial}}$.}
    \label{fig:model_weights}
\end{figure}

Let us now have a closer look on the main-diagonal entries of the model confusion matrix (``self-identification weights'') over an increasing data set size in Figure \ref{fig:model_weights}.

It shows, that for the simplest model (SC) and clearly for the measurement data, perfect justification (model weight of $1.00$) is achieved very quickly. For the \textit{initial biofilm} model $M_\text{IB}$ and the \textit{full complexity} model $M_\text{FC}$, a larger data set size is required to justify their complexity. Since the weights for the more complex models do not stagnate at some point, we do not expect that a much larger data set is required to justify their complexity. 

When comparing both quantities of interest for the same data set size, the data-generating process for the calcite content is always identified with less confidence (i.e. obtains a lower weight) than for calcium.

\subsubsection{How Similar are the Models?}
\label{sec:results_similar}

Now we will assess the similarities between the different models looking on the off-diagonal entries in Figure \ref{fig:post_matrices}. For a single data point, we can clearly see that the models ``confuse'' their predictions, as the off-diagonal weights are relatively high. When the \textit{initial biofilm} model $M_\text{IB}$ or the \textit{full complexity} model $M_\text{FC}$ are the data-generating process for the calcite content, the weights for the other models are even larger than the main-diagonal entry. For increasing data set size, the dissimilarities between the models become more significant, but only for the calcium concentration. In contrast, the model confusion remains for the calcium predictions, i.e. the current data set size does not yield a clearer distinction between the models. However, using the full data set, the model confusion decreases significantly, only the similarity between the \textit{initial biofilm} model $M_\text{IB}$ and the \textit{full complexity} model $M_\text{FC}$ remains clearly visible. For both calcite and calcium, $M_\text{IB}$ and $M_\text{FC}$ are similar, since they both have a relatively high weight, when the other one generated the data. Having a look only at the calcite content shows that even when the \textit{initial biofilm} model $M_\text{IB}$ is the data-generating process, the \textit{full complexity} model $M_\text{FC}$ obtains a higher weight, which means that the model cannot be justified with this data set size \citep{schoniger2015finding}.

\subsubsection{How Good Do the Models Fit the Data?}
\label{sec:results_fit}

\begin{figure}[hptb]
    \centering
    \includegraphics[width = 1\textwidth]{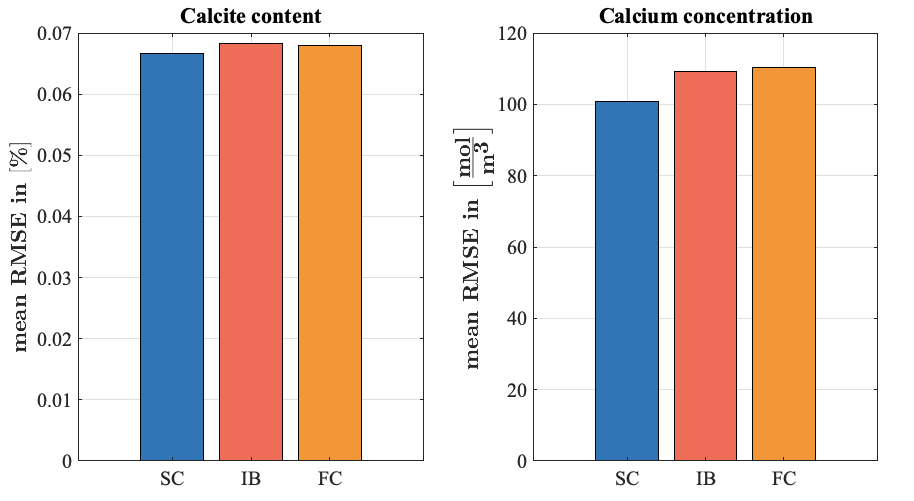}
    \captionof{figure}{Mean RMSE between the different model outputs and the measurement data.}
    \label{fig:rmse}
\end{figure}

In a last step, we will analyze the goodness-of-fit of the models to the measurement data. Figure \ref{fig:rmse} shows the root mean squared errors (RMSE) between the different model outputs and the measurement data, averaged over all model outputs evaluated on $P$ different collocation points:

\begin{align}
\sum\limits_{i=1}^{P} \sqrt{\sum\limits_{j=1}^{N_\text{s}} \left(M_{k,j}\left(\boldsymbol{\omega}^{(i)}\right) - \boldsymbol{y}_{0,j}\right)^2}\text{,}
\end{align}

with $\boldsymbol{y}_{0,j}$ the vector of measurements at position $j$ of total length $N_\text{s}$ and $M_{k,j}\left(\boldsymbol{\omega}^{(i)}\right)$ the model output of model $M_k$ at position $j$ evaluated at collocation point $\boldsymbol{\omega}^{(i)}$.
The RMSE values for different predictions of the same model (different evaluations on different collocation points) were averaged to obtain one representative value per model. For all models the mean RMSE is almost identical in comparison, but in both cases it is smallest for the \textit{simple chemistry} model $M_\text{SC}$ (the error for calcium is higher since the output its magnitude is much higher than for the calcite). With regard to the BMS analysis it shows that the small BMS weights of the \textit{initial biofilm} model $M_\text{IB}$ and the \textit{full complexity} model $M_\text{FC}$ stem from an only slightly better goodness-of-fit, while the models are much more complex than the \textit{simple chemistry} model $M_\text{SC}$. Remember that a more complex model needs to have a significantly better goodness-of-fit to justify its complexity \citep{schoniger2015finding} (and to achieve a similar weight as a simpler model). Furthermore, it is interesting that the weight of the \textit{initial biofilm} model $M_\text{IB}$ is smaller than the one for \textit{full complexity} model $M_\text{FC}$ for the same data set size, although the \textit{full complexity} model $M_\text{FC}$ is slightly more complex. Therefore, the high computational effort of the \textit{initial biofilm} model $M_\text{IB}$ is not justified.

\subsubsection{Conclusions}
\label{sec:results_conclusions}

Combining the insights from the Bayesian model justifiability analysis and the goodness-of-fit analysis, we draw the following conclusions about the \textit{initial biofilm} model $M_\text{IB}$ and \textit{simple chemistry} model $M_\text{SC}$ as simplifications of the \textit{full complexity} model $M_\text{FC}$.
The \textit{full complexity} model $M_\text{FC}$ provides moderate BME values in the BMS analysis and does not use its full potential according to the Bayesian model justifiability analysis. Additionally, $M_\text{FC}$ provides unsatisfactory goodness-of-fit to the measurement data and cannot capture the underlying physical process reasonably well. The \textit{simple chemistry} model $M_\text{SC}$ for calcite and calcium obtains the same weight of $1.00$ in the BMS analysis (Figure \ref{fig:model_weights}) and Bayesian model justifiability for (Figure \ref{fig:post_matrices}) with the full data set. Therefore, the \textit{simple chemistry} model $M_\text{SC}$ uses its full potential to represent the data and it captures the response of the underlying physical system appropriately.

\section{Summary and Conclusions}
\label{sec:summary}

Bayesian model selection (BMS) cannot only be used for ranking models based on their goodness-of-fit to measurement data and parsimony, but also to quantify similarities among models.
This work introduces surrogate-based Bayesian model justifiability analysis for analyzing microbially induced calcite precipitation models in porous media. The suggested framework offers a rigorous pathway to address so-called conceptual uncertainty, i.e. which model is best suited for describing the underlying physical system. The justifiability analysis compares the models among each other and the available measurement data. 

Applying the justifiability analysis in addition to the BMS analysis yields a better insight on why a model wins the BMS ranking: either because it really fits the measurement data best or only because the data set size is too small to identify a more complex model, that actually fits better. Thus, the current best model is only best in the case of the given too limited data set size \citep{schoniger2015finding}.

The BMS and justifiability analysis were performed using surrogate models, which were built via an arbitrary polynomial chaos expansion (aPC) in order to assure feasibility of the analyses for computationally demanding biogeochemical models. The aPC accelerates the analysis, which requires a large number of model evaluations, by reducing the required number of evaluations of the original model. We apply Bayesian iterative updating of the surrogate models improving their accuracy while incorporating measurement data. In order to account for the error, that arises by comparing the surrogates instead of the original models correction factors for the calculated weights were introduced. The correction factor proposed by \citet{mohammadi2018bayesian}, correcting the comparison of model and measurement data, was extended to a novel correction factor for a comparison between two models. It helps to perform reliable surrogate-based Bayesian model justifiability analysis.

Applying the introduced Bayesian justifiability analysis to three different models (\textit{simple chemistry} model $M_\text{SC}$,\textit{initial biofilm} model $M_\text{IB}$ and \textit{full complexity} model $M_\text{FC}$), we compare the models to measurement data and among each other. The comparison is based on the predictions of calcite content and calcium concentration at different data points in space and time. The justifiability analysis has shown that the \textit{simple chemistry} model $M_\text{SC}$ and the \textit{full complexity} model $M_\text{FC}$ for calcite and calcium and the \textit{initial biofilm} model $M_\text{IB}$ only for calcium identify themselves best, in comparison to the other models, when a certain data set size is used. The \textit{simple chemistry} model $M_\text{SC}$ even achieves perfect justification with a weight of $1.00$.

The analysis has also revealed that the data set size is too small for justification of the \textit{initial biofilm} model $M_\text{IB}$ in terms of the calcium concentration, since its diagonal entries of the model confusion matrix are always smaller than $0.5$. Further, it shows that the \textit{initial biofilm} model $M_\text{IB}$ and the \textit{full complexity} model $M_\text{FC}$ are similar in terms of both quantities of interest (calcium concentration and calcite content). Additionally, performing the conventional BMS analysis reveals the \textit{simple chemistry} model $M_\text{SC}$ as the best model in the model set, because of its best trade- off between goodness-of-fit to the measurement data and its sufficiently small degree of complexity.

The proposed analysis provides an extension of the very general justifiability analysis by \citet{schoniger2015finding} that makes it applicable for computationally expensive models. It can be concluded that the results for surrogate models followed the intuitively assumed preference for the simplest model when only little data is available. This makes the method ideal for application cases where the same situation, little data and computationally expensive models, appears. Although this method poses an effective way of comparing computationally expensive models their computational cost must not be disregarded. With increasing computational cost the number of model evaluations decrease for a given period of time, which leads to a more imprecise surrogate model and therefore less reliable results in the justifiability analysis.

\newpage

\begin{acknowledgements}
The authors would like to thank the German Research Foundation (DFG) for financial support of the project within the Collaborative Research Center 1253 CAMPOS (DFG, Grant Agreement SFB 1253/1 2017), Collaborative Research Center 1313 (SFB1313) (DFG, Project Number 327154368), DFG project number 380443677 and the Cluster of Excellence EXC 2075 ``Data-integrated Simulation Science (SimTech)'' at the University of Stuttgart under Germany's Excellence Strategy - EXC 2075 - 390740016.

Measurement data is available in \citet{hommel2015revised}, data for the MICP models and the justifiability analysis is available online in the repository https://git.iws.uni-stuttgart.de/dumux-pub/scheurer2019a.
\end{acknowledgements}

\begin{appendices}
\section{Computational details for the overdetermined system of equations \label{app:system}}
The solution of the overdetermined system needs to be approximated by minimizing the Euclidian norm ($L_2$ norm) of the residual:
\begin{align*}
    \min_{\boldsymbol{\omega}} ~ \|\boldsymbol{\Psi}(\boldsymbol{\omega}) \cdot \textbf{c}(\textbf{x},t) - \boldsymbol{M_k}(\textbf{x},t;\boldsymbol{\omega})\|_2 \text{.}
\end{align*}
via a linear regression:

\begin{align*}
    \boldsymbol{\Psi}^T(\boldsymbol{\omega}) \cdot \boldsymbol{\Psi}(\boldsymbol{\omega}) \cdot \textbf{c}(\textbf{x},t) = \boldsymbol{\Psi}^T(\boldsymbol{\omega}) \cdot \boldsymbol{M_k}(\textbf{x},t;\boldsymbol{\omega}) \text{.}
\end{align*}

The new system is determined again and can be solved with the help of the pseudoinverse:

\begin{align*}
    \textbf{c}(\textbf{x},t) &= \left(\boldsymbol{\Psi}^T(\boldsymbol{\omega}) \cdot \boldsymbol{\Psi}(\boldsymbol{\omega})\right)^{-1} \cdot \boldsymbol{\Psi}^T(\boldsymbol{\omega}) \cdot \boldsymbol{M_k}(\textbf{x},t;\boldsymbol{\omega}) 
    \label{eq:pseudo1}
    \\
    \textbf{c}(\textbf{x},t) &= \boldsymbol{\Psi}^+(\boldsymbol{\omega}) \cdot \boldsymbol{\Psi}^T(\boldsymbol{\omega}) \cdot \boldsymbol{M_k}(\textbf{x},t;\boldsymbol{\omega}) \text{,}
\end{align*}

where $\boldsymbol{\Psi}^+(\boldsymbol{\omega})$ denotes the pseudoinverse.
\end{appendices}

%
%


\bibliographystyle{spbasic}      

\bibliography{literature.bib}   


\end{document}